\documentclass{aa}
\usepackage{graphicx}

\begin{document}

\title{ Super-Eddington 
accretion rates in Narrow Line Seyfert 1 galaxies}

\author{Suzy Collin\inst{1}, Toshihiro Kawaguchi\inst{1,2}}

\offprints{Suzy Collin (suzy.collin@obspm.fr)}

\institute{$^1$LUTH, Observatoire de Paris, Section de
Meudon, F-92195 Meudon Cedex, France\\
$^2$  Postdoctoral Fellow of the Japan Society for the Promotion of 
Science}

\titlerunning{The accretion rate in NLS1s}
\authorrunning{Suzy Collin and Toshihiro Kawaguchi}

\abstract{We use the BH masses deduced from 
     the empirical relation  of Kaspi et 
al. (2000) between the size of the 
 Broad Line Region (BLR) of Active Galactic Nuclei (AGN) and the 
 optical luminosity, to compute their accretion rate in four 
     samples of AGN, assuming that the optical luminosity is provided 
     by the accretion disc. 
     We show that Narrow Line Seyfert Galaxies 1 (NLS1s) accrete
 at super-Eddington rates, while their luminosity stays of the order of 
 the Eddington limit. We take into account the 
 possibility of a non-viscous energy release inversely proportional to 
 the square of the distance in the gravitationally 
 unstable region of the disc emitting a fraction of the optical 
 luminosity. It 
 leads to a smaller accretion rate and to a redder continuum
  than a standard disc, which agrees better 
 with the observations. 
  The observed bolometric luminosities appear to saturate
	at a few times the Eddington luminosity for super-Eddington accretion 
	rates,
	as predicted by slim disc models. They favor a Kerr BH 
 rather than a Schwarzshild one.
 Even when the accretion 
 rate is super-Eddington, it stays always
 of the order of a few M$_{\odot}$/yr, irrespective of the BH mass, 
 indicating that the growing of the BH 
is mass supply limited  and therefore regulated by an exterior mechanism, 
and not Eddington limited. The mass of the BH increases 
 by one order of magnitude in a few 
  10$^7$ years, a time smaller than that necessary for changing the 
  bulge mass. This is in agreement with recent claims that the BHs of NLS1s 
   do not follow the same black hole - bulge relation as other galaxies. 
   Since they represent about 10$\%$ of AGN up 
   to a redshift of 0.5, these ``super-active" phases should play an 
   important role in shaping the mass function of local BHs. We 
   finally discuss the possibility that the masses could be systematically 
   underestimated due to an inclination effect, and we conclude that 
   it could indeed be the case, and that
    the accretion rates could thus be strongly overestimated in 
   a small 
   proportion of objects, possibly explaining the existence of 
   apparently extremely high accretors.

}

\maketitle

\section{Introduction and rationale}

The evolution of massive black 
holes (BHs) in relation with their host galaxy is presently intensively debated. 
Massive black holes
seem present in
 all galactic nuclei, independently of their level of activity. 
In about 40 inactive nearby galaxies, their mass was found 
proportional to the luminosity of the 
bulge of the host galaxy (Magorrian et 
al. 1988). Ferrarese \& Merritt (2000)
and Gebhardt et al. (2000a) showed that a tighter relation 
exists between the mass of the BH, $M$, and the dispersion 
velocity $\sigma_{\rm B}$ 
of the bulge. The slope of the relation is still debated,
and the recent work of Tremaine et al. (2003) gives a 
value  close to 4.  
 Several mechanisms accounting for 
this relation have been proposed (Silk \& Rees 1998,
 Umemura 2001, King 2003). When $\sigma_{\rm B}$ is expressed in terms 
of the bulge mass, it leads to 
 $M \sim 0.002 M({\rm 
Bulge})$. It is thus clear that the growth of the BH and 
the evolution of the host galaxy are related, so it is generally 
assumed that
 their co-evolution is 
mainly the result of 
merger events within the hierarchical scenario of large 
structure formation (Haehnelt, Natarajan \& Rees 1998, Kaufman \& Heahnelt 2000, 
Menou, Haiman \& Narayaman  
2001, Hatziminaoglou et al. 2003). 

However this scenario begins to be questioned seriously. It is indeed 
difficult to explain how smaller BHs
 grow at lower redshifts and more massive ones at higher redshift. So
Marconi et al. (2004) propose that local BHs  grow mainly
during Active Galactic Nuclei (AGN) phases.
This raises immediately the question
 whether BHs in local AGN and in quasars follow the same 
 BH/bulge
relationship as other galaxies. 

The BH masses in AGN are not 
determined like in inactive 
galaxies by the study of the stellar rotation curve close to the 
center. In about 40 AGN, they are determined directly through 
reverberation mapping (Wandel et al. 1999, Kaspi et al. 2000),
 which yields an empirical relation between the luminosity and 
the size of the 
Broad Emission Line Region (BLR), and then to the BH mass, using
the Full Width at Half Maximum (FWHM) of the broad lines  as a 
surrogate of their dispersion velocity and assuming 
that the BLR is gravitationally bound to the BH, an assumption 
confirmed by detailed studies (Peterson \& Wandel 1999 
and 2000). In the 
 other AGN the BH masses are determined indirectly 
assuming that the same relations hold.
 Wandel (1999) showed that Seyfert galaxies 
have lower BH to bulge mass ratios than inactive galaxies, but the revision of 
the Magorrian relation leads  to conclude finally 
that it is not the case (Laor 2001, Wandel 2002, Gebhardt et al. 2000b). 

However the status of Narrow Line Seyfert 1 galaxies (NLS1s) is not 
well established 
 in this context. NLS1s constitute about 10$\%$ of  
Seyfert nuclei and quasars up to a redshift of 0.5 (Williams, Pogge, 
\& Mathur 2002). Though they are known since a long time (Osterbrock \& Pogge 
1985), their nature is still not well understood. 
Besides the ``narrowness" of their broad lines, 
these galaxies share common properties, such as 
 strong FeII permitted lines and weak forbidden [OIII] lines, a strong X-ray 
 variability and a big soft X-ray hump
  (see several reviews in Boller et al. 2000). 
 Mathur, Kuraszkiewicz \& Czerny (2001)
 suggested that the BH/bulge 
mass ratio is smaller in NLS1s, and Wandel 
(2002) found that 
$M \sim 10^{-3}$ to $10^{-4} M({\rm 
Bulge})$, a smaller value than for broad line AGN (BLS1s).
 Both papers are based on a very limited sample, and 
are prone to statistical uncertainties. Moreover, in NLS1s the bulge
mass is generally not deduced from the stellar dispersion velocity
 but from the 
 width of the [OIII]5009 line assumed to be proportional to it,
  following a suggestion of Nelson and Whittle (1996) 
 for Seyfert 1 galaxies (actually Wandel (2002) used direct 
 measurements of the bulge luminosity). Wang and Lu (2001)
  argued that the [OIII] width is not accurately determined
   in NLS1s, owing to the weakness 
 of the line and to the presence of a blue wing, both effects leading
to overestimate $\sigma$([OIII]) and therefore the bulge mass. 
However Grupe \& 
 Mathur (2003) confirmed the previous result of
  Mathur et al. (2001) 
 with a complete X-ray selected sample of NLS1s, even when taking into account 
the presence of the blue wing of the [OIII] line, and she claims
  that NLS1s occupy distinct regions in the BH/bulge mass relation. Botte et 
al. (2004) do not confirm this result, and from a study of the 
photometric 
properties of the host galaxies they find that the NLS1 galaxies seem to share 
the same BH/bulge mass relation as ordinary Seyfert, and simply occupy the 
lower ranges of the $M - M({\rm 
Bulge})$ plane. Bian and Zhao (2003) came to an opposite conclusion,
based also on the bulge luminosity (we recall that the relation 
deduced from the bulge luminosity and the host properties is more 
dispersed than that deduced from the dispersion velocity), but found 
that NLS1s do not follow the ordinary relation when using the 
[OIII] line as an indicator of the dispersion velocity (Bian \& Zhao 
2004). Finally Botte et al. (2004) show that there is a smooth
relation between the BH mass vs. the bulge luminosity
 for different classes of AGN,
while there is a jump between the BH mass v.s. the [O III] width.
The latter finding is consistent with what was claimed by Grupe \& Mathur (03)
and by Bian \& Zhao (04).

One sees that the problem of the BH/bulge mass relation in NLS1s 
is presently highly controversial.  It has important
  cosmological consequences.
If BHs in NLS1s are undermassive with respect to their host bulge, 
it would imply
  that these galaxies are ``young", in the sense that they 
  are still in the
 process of building their BH. It would mean
that BHs and galaxies do not evolve concomitantly
 (Mathur 2000, and Grupe \& Mathur 2003). We will show here that there 
 is a strong reason to
 believe this is true, because NLS1s seem to be accreting at 
 super-Eddington rates and therefore the time scale for the growing 
 of their central black holes could be extremely short.
 \bigskip
 
It is now widely admitted that NLS1s are radiating close to the 
 Eddington luminosity $L_{\rm Edd}$. This result is simply obtained from the 
 mass-luminosity-FWHM 
 relations mentioned above. A few objects might even have  super-Eddington 
 bolometric luminosity, depending on the conversion factor used 
 to transform the
 optical-UV luminosity into a bolometric one, and on the adopted Hubble 
 constant, but it never exceeds a few $L_{\rm Edd}$.
  From this result many people assuming that the efficiency 
  factor for conversion of mass into energy is constant and of the 
  order of 0.1 deduce 
 that these 
 objects are also {\it accreting} close to their Eddington limit. 
 
 But why 
 would it have to be so? Super-Eddington accretion is 
indeed theoretically allowed.
Near the BH,
 the gas forms an accretion disc, which is supposed to emit the ``Big Blue 
 Bump" (BBB). The accretion rate and the BH mass determine the spectral 
 distribution and the flux of the BBB. It is thus possible to 
 determine the accretion rate when the mass is known. It was 
 performed by Collin et al. (2002, hereafter 
 referred as C02), using the sample of Kaspi 
et al. (2000) for which the BH masses are deduced from reverberation 
mapping, and assuming that the optical luminosity
 is provided 
 by a standard accretion disc (once the luminosity of 
the underlying galaxy 
has been subtracted). They found that a fraction of objects 
is accreting at super-Eddington rates, while their optical 
luminosity stays lower than or of the order of 
the Eddington luminosity. Actually, when
  the accretion rate is close 
 to, or larger than 
 the Eddington limit, 
accretion close to the BH does not proceed through a ``thin", 
but a ``slim" disc whose 
cooling time is larger than the viscous time, so 
energy is advected towards the BH before being radiated. The 
mass-energy conversion
efficiency $\eta$ thus decreases as the 
accretion rate increases, and the luminosity increases only 
logarithmically with the accretion rate (Abramowicz et al. 1988,
Wang et al. 1999, Fukue 2000,
Mineshige et al. 2000, Wang \& Netzer 2003, Kawaguchi 2003). 
The emission of such a disc is characterized by a soft X-ray bump as 
those observed in NLS1s.  
 Kawaguchi (2003), and Kawaguchi, Pierens \& Hur\'e (2004, hereafter 
 called KPH) have confirmed that the 
 overall Spectral Energy Distribution (SED) of the two most super-Eddington 
 accretors are well fitted by the 
 emission of a slim disc. Finally,  Wang (2003) noted
that super-Eddington accretion should lead to a limit relation 
between the BH mass and the FWHM of the lines, and he found several objects 
satisfying this 
relation, indicating that they radiate close to 
their Eddington luminosity, but accrete above the 
Eddington limit. 

\bigskip

There were only a few NLS1s in the Kaspi et al. sample studied in C02. 
Moreover the sample is not statistically complete since half of the objects 
are nearby Seyfert nuclei chosen mainly for their high degree 
of variability. The recent release of 
several complete samples including a large number of NLS1s, and the 
renewed interest for these objects since a few years, motivated 
us to conduct the same study on these new samples. While only 
standard discs were assumed in C02, here we take into account the 
deviation from the standard disc due to the disc self-gravity, which is 
particularly important in super-Eddington objects (cf. KPH). We use 
also the slim disc model to compute the bolometric luminosity as a 
function of the accretion rate.
We finally discuss some observational consequences not 
envisioned in C02. The model can 
account for 
the fact that the optical-UV continuum of NLS1s is redder than that of ordinary 
Seyferts (Constantin \& Shields 2003). The 
variation of the bolometric luminosity with the accretion rate agrees 
with the slim disc model. It explains why the FWHMs of 
the broad lines are larger than 700km/s.

In this paper, we only want to show some general trends and draw
qualitative conclusions concerning the accretion rates of NLS1s, using 
rough theoretical models of accretion discs and applying them to  
entire samples.

 Finally we insist on the fact that all along this paper we accept the commonly 
admitted statement that the narrowness of the lines of NLS1s is 
not due to an 
inclination effect, i.e. that NLS1s do not constitute a sample of 
normal Seyfert 1 nuclei whose broad line 
region is a rotating disc seen almost face-on. In this case, it 
is clear that the masses derived from the reverberation mapping 
formulae would be strongly underestimated, and consequently their luminosity
 (in terms 
of Eddington luminosity) overestimated. 

\bigskip 

In the following section, 
 we recall first how BH masses are determined and we 
present the samples. We 
 discuss 
the explanation of the
empirical relation between the luminosity and the size of the BLR.  
In Section 3, we summarize the theoretical model. Section 4
 is devoted to a discussion of 
the results, and in the last section we discuss the alternate 
 possibility that the masses of NLS1s could be underestimated and the 
   accretion rates overestimated.

\section{Determination of the BH masses}

\begin{figure}
 \centering
 \includegraphics[width=9cm]{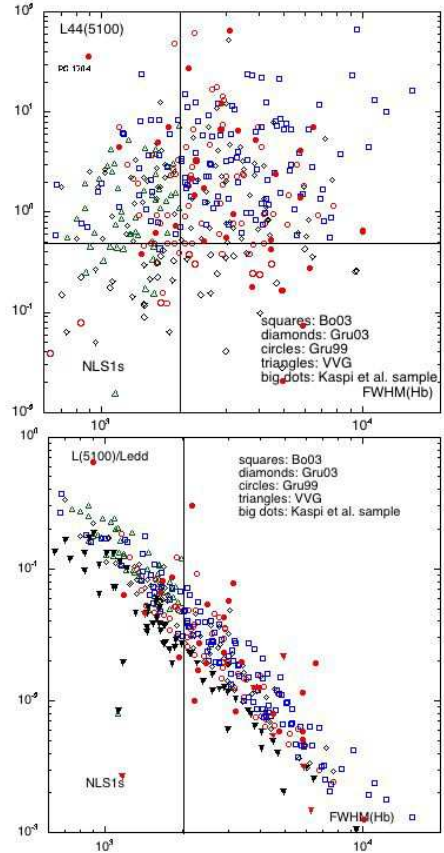}
\caption{Respectively L(5100) (top) and L(5100)/$L_{\rm Edd}$ (bottom) 
versus the FWHM for all samples. The black (respt. red) triangles indicate the 
 objects with L(5100) $\le 0.5\ 10^{44}$ ergs/sec for all samples 
 (respt. the Kaspi et al. sample). The object lying 
 much
 below the others is NGC 4051. }
 \label{fig-L5100-vs-FWHM}
\end{figure}

\subsection{The empirical mass-luminosity relation}

Reverberation mapping studies allowed to determine the size of 
the BLR in about 40 objects. It lead to the discovery of a
correlation between the radius of the region emitting the H$\beta$ line, 
which we will call $R{\rm (BLR)}$,
 and the monochromatic luminosity at 5100\AA, ${\rm L(5100)}=\nu 
 L_{\nu}(5100)$ (Kaspi et al. 2000):
\begin{equation}
R{\rm (BLR)}=32.9\times {\rm L(5100)}_{44}^{0.7}\ \ \ {\rm lt\ days},
\label{eq-RBLR-L5100}
\end{equation}
where ${\rm L(5100)}_{44}$ is expressed in 
10$^{44}$ 
erg/s. Though there is some uncertainty in the functional form 
of the relation (cf. Laor 2003, Netzer 2003), all recent papers adopt this relation 
to compute $R{\rm (BLR)}$ 
in quasars and Seyfert 
galaxies, when it has not been determined by reverberation 
mapping. 

It is now 
 well demonstrated that the broad H$\beta$ emitting region is 
 gravitationally bound to the BH (Peterson \& Wandel 2000).  
This gives another relation,
$M_{\rm BH} = R{\rm (BLR)}V^{2}/G$, where $G$ is the gravitational constant.
$V$ is generally taken equal to
$\sqrt{3}/2\ {\rm FWHM}$, corresponding to BLR clouds in random
orbital motion. The relation becomes, using Eq. \ref{eq-RBLR-L5100}:
\begin{equation}
 M_{\rm BH}= 5.8\ 10^{5} \times R{\rm (BLR/lt\ days)}
 {\rm \times (FWHM)}_{2000}^{2}\ {\rm M_{\odot}}, 
\label{eq-MBH-RBLR}
\end{equation}
where ${\rm (FWHM)}_{2000}$ is  the FWHM of the H$\beta$ line
expressed in 2000 km/s (we choose this value since 
NLS1s are defined by FWHM$\le 2000$ km/s).
Using Eqs. \ref {eq-RBLR-L5100} and \ref {eq-MBH-RBLR}, one gets a relation 
between $ M_{\rm BH}$ and ${\rm L(5100)}$ which allows to 
determine $ M_{\rm BH}$ as a function of the optical luminosity and 
the FWHM, without the need to know the size of the BLR. We stress 
however that the use of the FWHM as a surrogate of the dispersion 
velocity can lead to a systematic underestimation of the mass,
 if the BLR is a relatively flattened structure 
dominated by rotation, in which case the inclination of the system 
would play an important role (see Section 5).

\bigskip

These relations have important consequences. If one assumes 
that $L_{\rm bol}\sim 10\ {\rm L(5100)}$, a canonical value for the quasar 
continuum (cf. Elvis 1994, Laor et al. 1997), one gets from Eqs. \ref{eq-RBLR-L5100}
and \ref{eq-MBH-RBLR}:
\begin{eqnarray}
R_{\rm Edd}= && 0.35 {\rm L(5100)}_{44}^{0.3} {\rm (FWHM)}_{2000}^{-2} 
\\
\nonumber 
&& = 0.28 M_{7}^{0.43}{\rm (FWHM)}_{2000}^{-2.86} \ \ {\rm lt\ days},
\label{eq-Redd}
\end{eqnarray}
where we call $R_{\rm Edd}$ the Eddington ratio, i.e. the ratio of the
 bolometric upon the Eddington luminosity $L_{\rm Edd}=1.5\ 10^{45}M_{7}$,
  and $M_{7}$ the BH mass expressed in 10$^7$ M$_{\odot}$. 
It 
is obvious from this relation that NLS1s have larger Eddington ratios 
than BLS1s for a given BH mass.

\subsection {Comments on the luminosity-size relation}

 There are several possible explanations for this relation.
 Line emission 
can be suppressed by dust beyond the radius of sublimation, which 
corresponds to a given heating flux $\propto L_{\rm bol}/R^{2}$
(Netzer \& Laor
1993). But this constraint provides only an outer boundary
of the BLR.
 Nicastro (2000) proposed that clouds
are formed in a wind above the disc, close to the transition region 
between the gas and the radiation pressure dominated zones of the 
disc. However the size of the BLR 
depends both on the BH mass and on the luminosity, while the observations 
give only a luminosity dependence. 
 The striking similarity of AGN spectra led also to the idea that
  the ``ionization parameter"  
(i.e. the radiation pressure to gas pressure ratio 
or the photon density to gas density ratio, $\propto L_{\rm bol}/(nR^{2})$, 
$n$ being the electron number density) is constant among all objects. 
Actually the size-luminosity relation rather implies that the product 
of the density with the ionization parameter is constant. This is 
consistent with the so-called ``LOC" model.
 
In 1995, Baldwin et al. proposed
that the observed spectrum  of AGN is simply a consequence of the ability
of a photoionized medium to reprocess the underlying continuum ``as long as 
there are enough clouds at the correct
radius and with the correct gas density to efficiently form a
given line". In this
 ``Locally Optimally Emitting 
Clouds" (or LOC) model, each line 
is emitted preferentially at an appropriate ionizing flux $L/4\pi R^{2}$ 
corresponding to a given distance from the source \footnote{ 
This is actually closely related with 
the old idea of line saturation due to thermal 
quenching (Ferland \& Rees 1988, Collin-Souffrin \& Dumont 1989)}.  
According to the grid of photoionized models published by
 Korista et al. (1997) 
the ``optimal" ionizing flux 
$F_{\rm optimal}$ for the  
the H$\beta$ line does almost not depend on the density and on the spectral 
distribution of the 
ionizing continuum. It is of the order of $ 10^{8}$ erg s$^{-1}$ 
cm$^{-2}$.
This means that as long as there are clouds in a large range of 
radius  with the appropriate density
 (i.e. between 10$^{9}$ and 10$^{14}$ cm$^{-3}$) the ionizing continuum 
 will be reprocessed in the H$\beta$ line with a maximum efficiency 
 at an   
optimal distance 
$R_{\rm optimal}\sim 2\ 10^{17}L_{\rm ion,44}^{0.5}$ cm, where $L_{\rm ion}$
is the ionizing luminosity. 
  From 
 the Gru03 sample, one gets
L(5100)$\sim 0.1\ L_{\rm bol}^{0.7}$ (precisely  
L(5100)$=0.21\ L_{\rm bol}^{0.6}$ and $L_{\rm bol}= 17$ L(5100)$^{1.13}$, 
with a correlation factor of 0.9). Thus the observed relation transforms into 
$R_{\rm BLR}\sim {\rm a\ few}\ 10^{17}L_{\rm ion,44}^{0.5}$ cm, which 
is similar to the relation  expected for the LOC 
model (the ionizing luminosity being
 slightly smaller than the bolometric
luminosity). 

So  {\it the only necessary condition for the 
observed relationship is 
 the existence of clouds within a broad range of density 
at a radius smaller than the typical distance of the BLR}, say 
10$^{4}R_{\rm G}$. 
Collin \& Hur\'e 
(2001) suggested that such clouds form above the 
gravitationally unstable region of the disc. Since the disc becomes gravitationally unstable 
 at small radii compared with the size of the BLR (cf. later), 
 this condition is satisfied.  The BLR clouds would 
thus constitute simply the outer part of the region emitting the optical 
continuum. 
 Laor (2003) objected to this idea that ``since all
accretion discs must become gravitationally unstable far
enough from the center, this mechanism does not provide a
natural explanation for the apparent absence of a BLR in
some Agn". But there are actually several possible explanations for 
the absence of BLR. For instance, in low luminosity objects, it can be due to the 
suppression of the ionizing radiation in an Advection Dominated 
Accretion Flow (ADAF). It can also simply be caused by the absence of 
adequate  
physical conditions in the gravitationally unstable disc, like
 a too high or to small density.

\subsection{The samples}

\begin{figure*}
 \centering
 \includegraphics[width=18cm]{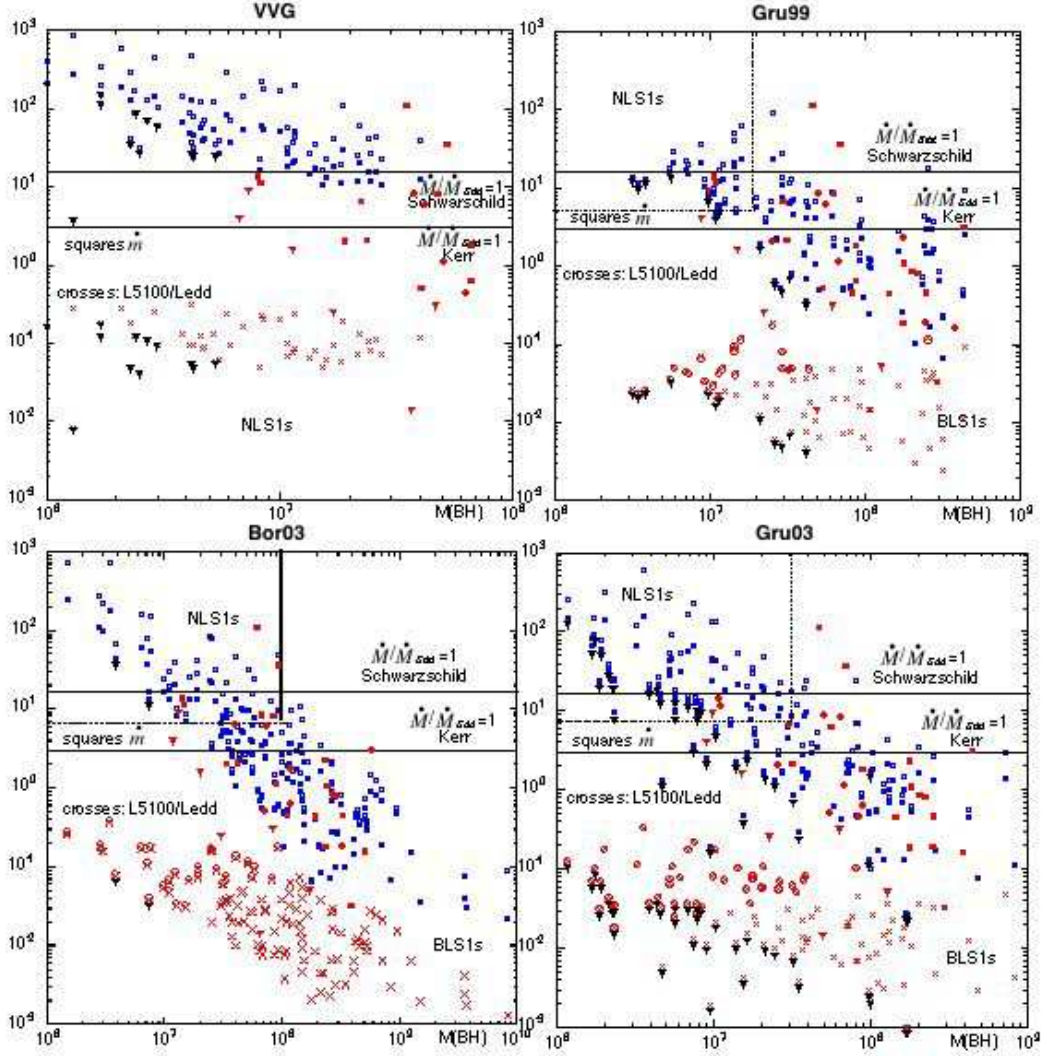}
\caption{ These figures display $\dot{m}$ 
 as a function of the BH masses for the four samples. The squares 
 give $\dot{m}$ 
 computed according both to the standard disc model (open squares),
  and to the self-gravitating 
 disc model with a
  viscosity parameter 
 $\alpha=0.01$ (filled squares). The 
 crosses give  L(5100)/$L_{\rm edd}$, and the crosses with open circles mark 
 the NLS1s. The two thick solid lines 
delineate the position of $\dot{m}$ for the NLS1s. The two horizontal lines 
 correspond to $\dot{M}/\dot{M}_{\rm Edd}=1$, where $\dot{M}_{\rm 
 Edd}={L}_{\rm Edd}/(\eta c^{2})$, in the case of a 
 Schwarzschild BH and of a Kerr BH. The black (respt. red) triangles indicate the 
 objects with L(5100) $\le 0.5\ 10^{44}$ ergs/sec (respt. for the 
 Kaspi et al. sample). Note that the VVG 
 sample consists only of NLS1s, thus circle symbols are not shown. }
 \label{fig-mdot-vs-M}
\end{figure*}

We use two complete samples including both NLS1s and BLS1s.

The recent data release of the Sloan Digital Sky Survey (SDSS) allowed 
Boroson to build an homogeneous sample of 107 low-redshift radio-quiet 
QSOs and Seyfert 1 galaxies (Boroson 2003). It is aimed at 
comparing the BH masses determined from the empirical 
relations with those deduced from the dispersion velocity of the [OIII] line,
 used as 
a surrogate of the stellar velocity dispersion. About one third of 
objects are NLS1s in this sample. It allows to study a 
large range of masses and luminosities. For each object the 
redshift, the FWHM(H$\beta$) and the BH mass are given, and we deduce 
the optical luminosity 
at emission
from Eqs. \ref{eq-RBLR-L5100}
and \ref{eq-MBH-RBLR}. We call this sample Bor03.

The second one is a complete sample of X-ray selected AGN (Grupe 
et al. 2003). According to the selection procedure, about half of 
the objects are NLS1s. L(5100) is given, but for an empty universe, so 
we made the conversion to $q_{0}=0.5$. We call this sample Gru03. 
It is particularly interesting for us as it gives an estimate 
of the bolometric luminosity of the objects based on the observed spectral 
energy distributions, which we will be able to 
compare with our models.  

We use also two other heterogeneous samples. 
Wang \& Lu (2001) deduced L(5100) from the B-magnitude using the V\'eron-Cetty et al.
 (2001)  sample, which contains 59 NLS1s, and they estimated the BH 
 masses using the previous empirical relations. After rejection of a 
 few objects for which the FWHM are controversial, 
 the sample was reduced to 54 NLS1s. We call it the VVG 
 sample.
 We also used an heterogeneous sample of soft X-ray selected AGN (Grupe et al. 
 1998, 1999), which 
 has the advantage of giving optical indices
 useful to check our models. We also made the conversion from $q_{0}=0$ to $q_{0}=0.5$.
 We call it Gru99. Note that a few objects 
 are also in Gru03.
 
 The samples have not been corrected for the stellar contribution of the host 
 galaxy to the optical luminosity. It is certainly important for low luminosity 
 AGN, but not when the optical luminosity is larger than a few 10$^{43}$ 
 ergs/sec.  {\it In the 
 following we will distinguish or suppress all these weak objects from the samples, 
 so we can be fairly confident that the results will not be contaminated by 
 the host galaxy}.

 Fig. \ref{fig-L5100-vs-FWHM} displays respectively L(5100) (top) and 
 L(5100)/$L_{\rm Edd}$ 
 (bottom) 
versus the FWHM for all samples. We note immediately the strong 
difference between these two graphs. While the first one shows a very 
loose correlation, corresponding to the absence of low luminosity 
objects
with large FWHMs and of high luminosity objects with small FWHMs, the 
second one shows a tight correlation with a slope equal to -2, which
 is expected according to the first line of Eq. 3.
The black triangles indicate the 
 objects with L(5100) $\le 0.5\ 10^{44}$ ergs/sec: note 
that these low luminosity objects share the same relation as the others.

 In an aim of comparison, we have added on these figures the 
objects where the BH masses have been determined directly by 
reverberation mapping (we call these objects the ``Kaspi et al. sample", though 
half of them were not observed by Kaspi et al. 2000). They 
span the same range of luminosities as the other samples. But first, they 
show a looser correlation between L(5100)/$L_{\rm Edd}$ and the FWHMs; 
it is expected as the determination of the mass in the other 
objects makes use of an exact relation $L-R$(BLR), not taking into account its 
error bars. And second, the relation should be extrapolated to 
 values of the mass and of the 
Eddington ratio smaller by a factor of 5. 
It should be kept in mind in the following analysis. 
Note that the values of the luminosities used in this figure 
correspond to $H_{0}=75$ km/sec/Mpc, while CO2 assumed $H_{0}=50$ km/sec/Mpc.

\section{The accretion disc model}

\begin{figure}
 \centering
 \includegraphics[width=9cm]{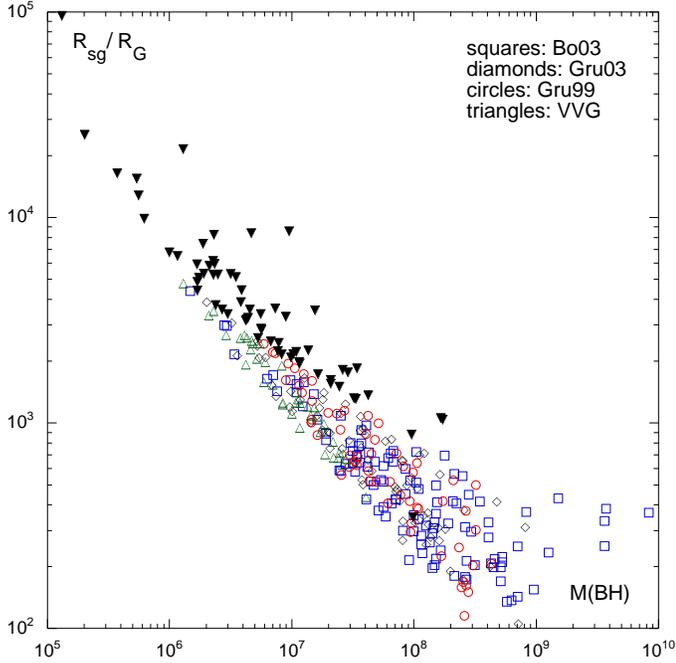} 
\caption{$R_{\rm sg}/R_{\rm G}$ versus $M$ 
   for all samples, for the self-gravitating disc with $\alpha=0.01$.
    The black triangles indicate the 
 objects with L(5100) $\le 0.5\ 10^{44}$ ergs/sec for all samples. }
 \label{fig-RsgsRg-Tsg01} 
\end{figure}

Since more than fifteen years it is widely admitted that 
the ``infrared bump" at a few microns and the ``Big Blue Bump" observed in radio quiet 
quasars and Seyfert nuclei are both due to thermal emission, respectively by 
hot dust heated by the UV-X continuum, 
and by the accretion disc (Sanders et al. 1989). In this picture, the observed ``dip" at 
$\sim $5000\AA$\ $ 
 in the log$(\nu F_{\nu})$ versus log$\nu$ 
curve corresponds to the junction between these two processes, the hot
dust close to the sublimation temperature (1700K) being unable to radiate 
appreciably below 1$\mu$m. In particular the idea of an underlying 
non-thermal power 
law continuum which was invoked in the past and used to model 
the infrared to UV emission of AGN has been completely left over. So 
{\it the emission
 at 5000\AA$\ $ should be due 
entirely to the accretion disc}, unless another medium can give rise to 
a smooth featureless optical continuum. 
The problem was discussed in C02, and they showed that
 it would require the existence of a very dense, optically 
thick and relatively cold medium. It is difficult to find for 
such a medium another location than an optically thick accretion disc.

 For a ``standard" thin Keplerian disc where 
 gravitational energy is released locally through turbulent viscosity,
  the effective 
 temperature $T_{\rm eff}$ at a 
distance $R$ from a BH of mass $M$ is:
\begin{equation}
\sigma T_{\rm eff}^4 = {3GM\dot{M}\over 8\pi R^3} f(R),
\label{eq-dissipation}
\end{equation}
where the non-dimensional factor $f(R)$ takes into account the boundary 
conditions, and is equal to unity at large radii (cf. for instance the 
book of 
Frank, King \& Raine 2002). 

Each spectral band is emitted around a given radius, and the optical band 
corresponds to a large distance from the black hole, typically 10$^{3}R_{\rm G}$ 
($R_{\rm G}$ being the gravitational radius $GM/c^{2}$). At such large 
radii, the disc is dense, relatively cold and optically thick, and
its local emission spectrum is close to a black body
 at the temperature $ T_{\rm eff}$ (cf. Collin 2001; note that it is 
 not the case 
at smaller radii, i.e. in the EUV band). Integrating over the disc the 
Planck law with $T\propto R^{-3/4}$, one
finds for the luminosity at a frequency $\nu$:
\begin{equation}
\nu F_{\nu}= {8\pi^{2}  h \nu^4 \over 
c^2 }
\int_{\rm Rin}^{\rm Rout} {RdR \over exp(h\nu/kT)-1}\ \propto \nu^{4/3},
\label{eq-Lnu}
\end{equation}
where ${\rm Rin}$ (respt. ${\rm Rout}$) is the inner 
 (respt. the outer)
 radius of the accretion disc.  
 
 So it is possible, using Eqs. \ref{eq-dissipation}  and 
 \ref{eq-Lnu}, to deduce the 
 accretion rate when the mass is known. One sees also from 
 these equations that L(5100) is approximately proportional to  
$(M\dot{M})^{2/3}$. This is not valid for very large masses and small accretion 
 rates, where $kT({\rm 
 Rin}$) is of the order of $h\nu_{\rm opt}$, or for truncated discs. 

For super-Eddington accretion rate this picture is changed.
The radiative efficiency per unit mass accretion
is expected to decrease due to the onset of photon trapping
(Begelman 1978).
As a result,
the emergent luminosity
from an accretion flow starts to saturate at a few
times $L_{\rm Edd}$ (Abramowicz 1988). 
 Self-similar solutions with super-Eddington accretion
rates (Fukue 2000; Wang \& Netzer 2003) are only valid inside the photon 
trapping radius, where
soft X-ray photons are emitted. However, full integration of
differential equations from far outside the photon trapping radius to
the vicinity of the central BH (Shimura \& Manmoto 2003; Kawaguchi 2003) 
is necessary in order to discuss
the broad-band spectra of NLS1s. We use the slim 
disc model for a Schwarzshild BH computed as in
Kawaguchi (2003), which is based on the code developed by
Matsumoto et al. (1984).
The effect of electron scattering (both in opacity and Comptonization) and 
relativistic correction
are included.
We take the viscosity parameter $\alpha$ equal to 0.1.
 Note that the slim disc is used here only to compute the bolometric luminosity.

Even if the accretion rate is very high (in Eddington value)
the optical luminosity is still emitted at a large radius 
where the accretion flow is not influenced by advection and photon 
trapping, except in the case of very high accretion rates ($\dot{M} 
\ge 3\ 10^{3}L_{\rm Edd}/c^{2}$, cf. KPH), and the standard disc model is 
valid. 
 The only deviation to the local blackbody in the 
optical region is due to 
electron scattering  (as modified blackbody, see Czerny \& Elvis 1987),
 which distorts the spectrum for super-Eddington accretion rates.
It is negligible as far as viscosity is small 
($\alpha \le 0.1$) and the BH mass is small ($M \le 
10^{7}$M$_{\odot}$), so the distortion is not very important for 
NLS1s (cf. KPH), and we 
will neglect it in this paper.

However an important fact should not be forgotten, which acts also 
for modest accretion rates but is very 
important for super-Eddington accretion rates. 

At about the distance 
of the optical emitting region, 
the disc becomes
self-gravitating, i.e. the vertical component of the BH gravity 
becomes smaller than the disc's own gravity. This occurs 
beyond a critical radius $R_{\rm sg}$ corresponding to a density: 
\begin{equation}
 \rho_{\rm sg}={\Omega_{\rm K}^2\over 4\pi G}
\label{eq-zeta}
\end{equation}
where  $\Omega_{\rm K}$ is the keplerian velocity.
The disc 
is then locally gravitationally unstable (Goldreich \& Lynden-Bell 1965).
 At radii larger 
than $R_{\rm sg}$, the structure of the disc is completely 
unknown. It could break into fragments, which can collapse and 
even form stars, or 
it can stay at the marginal instability limit if it can be sustained 
by some extra 
heating mechanism. In all cases the 
 region emitting the 5100 \AA$\ $ flux stays optically thick, and the local black body 
assumption is valid (cf. Collin \& Hur\'e 1999). 

  Koratkar \& Blaes (1999) stressed that the 
  standard disc 
model leads to a continuum bluer than 
 the average AGN continuum, which has a mean spectral index of 0.3 to 
 0.5 (we define the spectral $\alpha_{\rm opt}$ as $L_{\nu}\propto 
\nu^{-\alpha_{\rm opt}}$). Actually it is a problem only for small BH masses and 
large accretion rates. In the case of large BH masses and small 
accretion rates,
the optical spectrum-UV spectrum is emitted by the 
Wien part of the Planck function, and is redder than $\nu^{1/3}$.

Several   
sources of heating can overcome the gravitational viscous 
release
 beyond the self-gravitational radius. The disc can be irradiated by 
the central source of UV-X continuum
 if it is ``flaring" (i.e. if its 
thickness varies more rapidly than the radius).  It can be
heated by gravitational instabilities (Lodato \& Bertin 2003),
 by the collisions of 
clumps (Krolik \& Begelman 1988), or by
 embedded stars (Collin \& Zahn 1999), and cer. In all cases,
 $ T_{\rm eff}$ will decrease less 
rapidly with increasing $R$ than in a ``standard" disc, and the 
observed continuum will be redder.  
For instance
Soria \& Puchnarewicz (2002) fit the spectrum of the NLS1 1 RE 
J1034+396 (this object is included in the following 
computations) by an irradiated accretion disc whose scale height to the 
radius  $H/R$ ratio increases rapidly with $R$, $ T_{\rm eff}$ being 
thus
proportional to $R^{-1/2}$. C02 have shown that in this case,
 in order to get a smooth optical
 continuum without an 
intense Balmer discontinuity, the density and the optical thickness of 
the irradiated 
medium should be very large.  This is impossible with a strongly flaring disc;
 a warped thin
disc would be a more appropriate solution. As we explained 
previously, such a disc would be gravitationally unstable at the 
distance of the region emitting the optical luminosity, and most 
likely very different from a standard one. In the case of heating by 
embedded stars, a very large number of massive stars would be 
necessary to account for the whole optical luminosity (Collin \& Zahn 
1999).

Since  the status of the 
unstable part of the disc is not known, we parametrize 
these effects by assuming that 
the energy release is proportional to 
$R^{-\beta}$, with $\beta$  smaller than 3 in the self-gravitating region.
 In this paper we will
assume the extreme case $\beta=2$: 
it corresponds to 
 $L_{\nu}\propto \nu^{-\gamma }$, with $\gamma =1/2$. In the 
 following computations this value is used into Eq. 5 instead of Eq. 4
  for $R \ge R_{\rm sg}$, 
 with the 
 continuity of the energy release at $R_{\rm sg}$.
Doing this we obtain an optical spectral index between -0.3 (corresponding to 
 the standard disc) and +0.5, 
depending on the proportion of the disc which is self-gravitating. It 
 is closer to the observed AGN 
 continuum. The effect on the bolometric luminosity of this additional energy release is 
 negligible, but
it increases the computed emission in the 
 optical and near-infrared spectral bands, and therefore  decreases
 the accretion rate necessary to account for a given optical 
 luminosity. 
   $R_{\rm sg}$
  is small for small values of $\alpha $. We have thus chosen a 
  relatively 
   small value of the viscosity parameter (0.01) in order to 
   underestimate  $R_{\rm sg}$, and therefore to {\it underestimate 
   also the accretion rate} with respect to a standard disc.
 
  However, we have to take into account the fact that the accretion 
disc cannot extend too much in the self-gravitating region, unless a 
mechanism can act to limit the disc density at exactly the marginal 
instability. Since we 
will see below that the self-gravitation radius is always smaller than 
10$^{4}R_{\rm g}$, we have decided in the following 
to limit the radius of the accretion disc at a 
value of 10$^5R_{\rm g}$. It is an arbitrary value, but we have no way to 
estimate the real extension of the accretion disc. Note that the 
dimension of the BLR is at most
 of this order in NLS1s, and it is difficult to accept the idea 
that the disc extends much further out. 
Note that for such a radius,
  the gravity of the galaxy does not dominate on the BH.

 If the disk is not self-gravitating and extends further out, it does 
not influence the optical emission. Indeed in this case
 one finds that $\lambda (10^5R_{\rm g})\sim
  20 M_7^{1/4} \dot{m}^{-1/4}\ \mu$m, which insures that the optical emission 
  is entirely produced inside $10^{5} R_{\rm g}$.
On the contrary, if the disk extends only up to $10^3R_{\rm g}$ or
 $10^4R_{\rm g}$, the computed optical emission  
would be smaller than for $R_{\rm out}=10^5R_{\rm g}$, 
and {\it the accretion rate would therefore be larger}.  
 
As an accretion disc with a super-Eddington accretion rate behaves 
like a
standard disc outside the photon trapping radius (KPH), we compute $R_{\rm sg}$ with the same 
analytical approximation as KPH, which gives expressions similar to 
the previous  
 detailed computations of Hur\'e (1998):
 \begin{equation}
 R_{\rm sg}=\left(R_{\rm sg,a}^{3}+R_{\rm sg,b}^{3}+R_{\rm 
 sg,c}^{3}\right)^{1/3}
\label{eq-rsg}
\end{equation}
where $R_{\rm sg,a}$, $R_{\rm sg,b}$, and $R_{\rm sg,c}$
are the self-gravitation radius in respectively the inner region 
dominated by radiation pressure and Thomson opacity, the intermediate 
region, dominated by gas pressure and Thomson opacity, the outer 
region dominated by gas pressure and atomic opacity:
\begin{eqnarray}
&&R_{\rm sg,a}=500 \left({\alpha \over 0.1}\right)^{2/9} 
M_7^{-2/9}\left({\dot{M}\over L_{\rm Edd}/c^2}\right)^{4/9} \ R_{\rm G}\\
\nonumber
&&R_{\rm sg,b}=11400 \left({\alpha \over 0.1}\right)^{14/27} 
M_7^{-26/27}\left({\dot{M}\over L_{\rm Edd}/c^2}\right)^{-8/27} \ R_{\rm G}\\
\nonumber
&&R_{\rm sg,c}=13400 \left({\alpha \over 0.1}\right)^{28/45} 
M_7^{-52/45}\left({\dot{M}\over L_{\rm Edd}/c^2}\right)^{-22/45} \ R_{\rm G}.
\label{eq-rsg123}
\end{eqnarray}
These expressions depend on the viscosity parameter $\alpha$. We will 
use $\alpha=0.3$, $\alpha=0.1$, and $\alpha=0.01$. A smaller value 
of $\alpha$ has a more 
profound influence on the disc structure, as it corresponds to a 
denser standard disc, and therefore a smaller value of $R_{\rm sg}$.

Let us now discuss the consequences of 
these relations in an approximate way. As we shall see later, none of the free 
parameters have a strong influence on the computed accretion rate, 
the main quantity that we want to determine. We have seen 
that for a standard disc, L(5100)$\  
\propto (M\dot{M})^{2/3}$. Using this relation, and Eqs.
\ref{eq-RBLR-L5100} and \ref{eq-MBH-RBLR}, we get:
 \begin{equation}
\dot{m} \propto {\rm FWHM}^{-4.28} M^{0.14},
\label{eq-mdot}
\end{equation}
where $\dot{m}$ is the accretion rate expressed in Eddington units, 
 $\dot{m}={\dot{M}\over L_{\rm Edd}/c^{2}}$. 
This is actually a very interesting result, which
 comes from the dependence of the size of the BLR on the 
luminosity and which shows that $\dot{m}$ 
depends almost only on the FWHMs, and very little on the BH mass. It 
is only approximate if the self-gravitating region of the disc is 
large. 
 It means that {\it  $\dot{m}$ can be deduced directly from the 
measurement of the 
FWHMs alone}.  
 
 \section{Results and discussion}
 
 We have applied our model to the samples, and we present now the 
 results. We use H$_0=75$ km/sec/Mpc, 
and $q_0=0.5$.  When the luminosities were given for another 
cosmological constant, we have made the conversion in the aim of 
uniformity.

We first draw the attention on a fact which is sometimes forgotten.
 Generally it is not the fluxes at Earth but the 
luminosities which are published in 
the literature, and they are computed assuming an isotropic 
emission. 
The monochromatic luminosity 
is thus equal to:
\begin{equation}
\nu_{\rm e}L(\nu_{\rm e})= 4\pi D^2 (1+z)^2 \times \nu_{\rm e} 
F_(\nu_{\rm e}) / {\rm 
Abs}(\nu_{\rm o}), 
\label{eq-L5100-obs}
\end{equation}
where $F$ is the flux observed at Earth, $\nu_{\rm e}$ (respt.$\nu_{\rm o}$) 
is the frequency at 
emission (respt. at Earth), $D$ is the proper distance 
of the object,
$z$ is the redshift, ${\rm Abs}(\nu_{\rm o})$
 the external
 (galactic) absorption. But an accretion disc does not emit 
 isotropically. The computed monochromatic luminosity given by Eq. 
 \ref{eq-Lnu}  or by its equivalent for the self-gravitating region
 should thus be multiplied by a factor 2cos($i$), where  $i$ the 
inclination of the disc axis on the line of sight, to be 
 identified with the published values. 
 
\begin{figure}
 \centering
 \includegraphics[width=9cm]{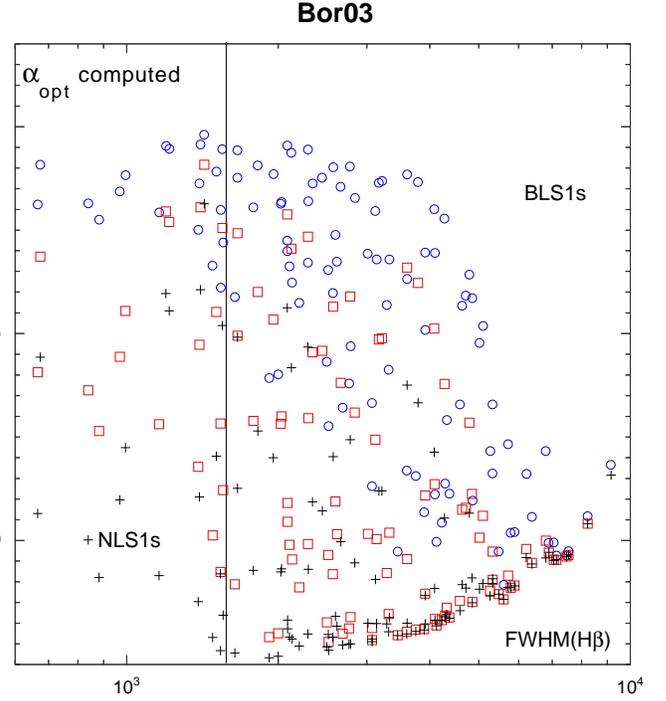}
\caption{The computed optical spectral 
index $\alpha_{\rm opt}$ measured between 4400 and 7000\AA (rest frame), for the 
Bor03 sample, excluding the 
 objects with L(5100)$\le 0.5\ 10^{44}$ erg/s. $\alpha_{\rm opt}$ 
 measured between 4400 and 7000\AA$\ $is computed with the 
 self-gravitating correction, for
  $\alpha$ equal respectively to 0.01 (blue circles) and 0.1 
 (red squares), and 0.3 (black crosses). We recall that $\alpha_{\rm 
 opt}=0.3$ for a standard disc. }
 \label{fig-bor-alfopt-vs-FWHM}
\end{figure}
 
\begin{figure}
 \centering
 \includegraphics[width=9cm]{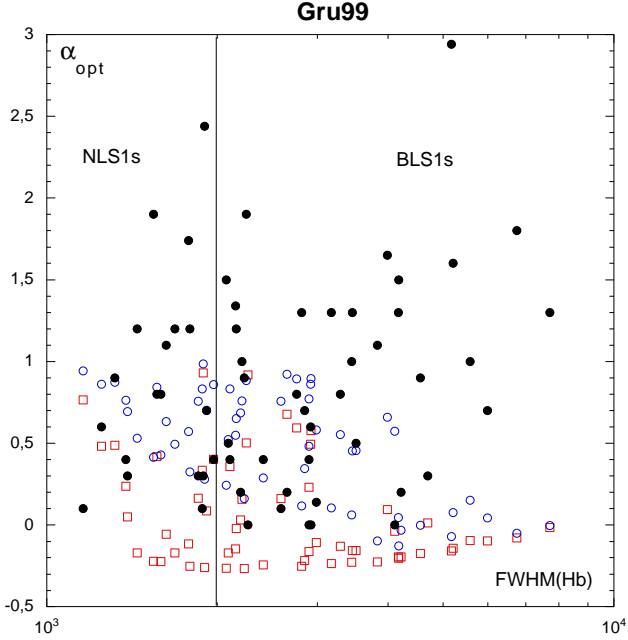}
\caption{ Comparison between the observed and computed optical spectral 
index $\alpha_{\rm opt}$, for the 
Gru99 sample, excluding the 
 objects with L(5100)$\le 0.5\ 10^{44}$ erg/s. The black filled circles 
 are the observed values, the open symbols are computed with the 
 self-gravitating correction, for
  $\alpha$ equal respectively to 0.01 (blue circles) and 0.1 
 (red squares). We recall that $\alpha_{\rm 
 opt}=0.3$ for a standard disc. }
 \label{fig-gru99-alfopt-vs-FWHM}
\end{figure}

  \subsection{Accretion rates}
 
 Figs. \ref{fig-mdot-vs-M} display $\dot{m}$ 
 as a function of the BH mass for the four samples. $\dot{m}$ is 
 computed according both to the standard disc model, and to the self-gravitating 
 disc model as explained in the previous section. In this latter 
 case, the results are shown for a
  viscosity parameter 
 $\alpha=0.01$. In all computations, 
 cos($i$) is set equal to 0.75. 
The objects with L(5100) $\le
0.5\ 10^{44}$ ergs/sec are indicated on the figures. There are only two 
such objects 
 (actually lying close to the limit) in Bor03. The Gru03 
 sample contains many low luminosity objects, but  
 a large number of NLS1s are above the luminosity limit. 
 
  We see 
 that the self-gravitation correction can 
decrease $\dot{m}$ by about a factor three for large values of $\dot{m}$, but 
has no influence on small $\dot{m}$.
 For larger values 
 of $\alpha$ and of $\gamma$, the difference between the standard and 
 the self-gravitating disc would be smaller.  So we can 
 consider that the two models here correspond  to a kind of ``error bar"
 on $\dot{m}$, for given BH mass and L(5100). Figs. \ref{fig-mdot-vs-M} shows also the 
 ``observed" ratio L(5100)/$L_{\rm Edd}$. We have noted the NLS1s, and 
 the 
 thick dotted lines 
 delineate the position of $\dot{m}$ for  NLS1s.  
 NLS1s
  always have BH masses smaller than 
 10$^{8}$ M$_{\odot}$, and they are located in the higher range of 
 L(5100)/$L_{\rm edd}$ and $\dot{m}$. It is interesting to note that 
 the four samples do not differ except for the range of masses and 
 luminosities,
 though they have been selected quite differently.  
 
  Again we added for comparison to these figures the results for the Kaspi et 
 al. sample, computed using only the standard disc emission (we recall 
 that the results differ from CO2 because we use here $H_{0}=75$
 instead of 50). As expected, the extrapolation by a factor 5 in mass range 
 of the empirical relationship 
 translates in an extrapolation of $\dot{m}$ by about a factor 30, as $\dot{m} 
 \propto \dot{M}/M_{\rm BH} \propto
L_{5100}^{3/2} \times M_{\rm BH}^{-2}$.

 Several other results appear on these figures. 
 
 First {\it $\dot{m}$ increases as the BH mass 
 decreases}. On the contrary, the  ratio L(5100)/$L_{\rm Edd}$ is always 
 smaller than 0.3, and seems about constant for the NLS1s. When 
 applying a standard correction $L{\rm bol}\sim 10\times {\rm L(5100)}$, one  
 concludes that {\it $L{\rm bol}$ saturates at about the Eddington luminosity, 
 whatever the BH mass}. This excludes the existence of the
large super-Eddington ratios proposed by Begelman (2002) due to 
the photon bubble instability. Thus, according to Eq. 3,
{\it there should be
a lower limit to the FWHMs of the order of 1000 $M_{7}^{0.15}$ km/s}
 unless the empirical relations 
do not apply to these objects. And indeed  FWHMs
 of the 
order of 100-500 km/s which would imply Eddington ratios larger 
than 10 have never been observed in Seyfert 1 nuclei.

 Second, the two horizontal lines 
 correspond to $\dot{M}/\dot{M}_{\rm Edd}=1$, where $\dot{M}_{\rm 
 Edd}={L}_{\rm Edd}/(\eta c^{2})$, in the case of a 
 Schwarschild BH ($\eta=0.057$) and of an extremely rotating Kerr BH 
 ($\eta=0.30$). We see that {\it the accretion rates of NLS1s 
 are always larger than the 
 Eddington rate in the case of Kerr 
 BHs, and mostly larger in the case of Schwarzschild BHs}. 
 
 \begin{figure}
 \centering
 \includegraphics[width=9cm]{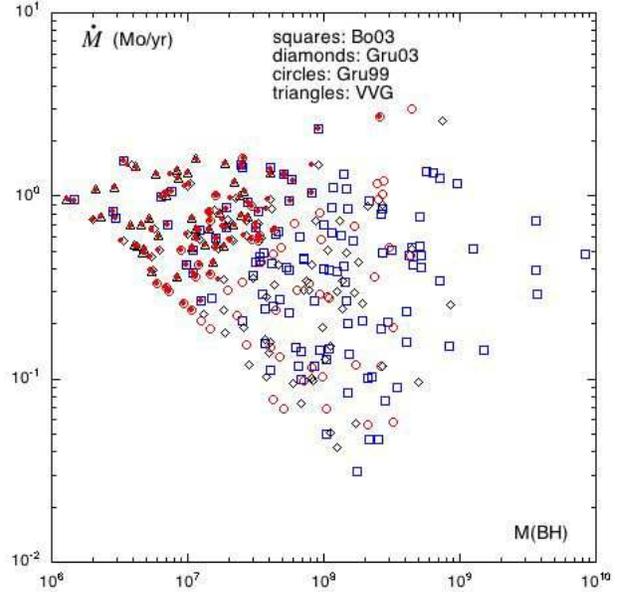}
\caption{Accretion rates in 
  M$_{\odot}$/yr for the four samples as a function of the BH masses, 
  excluding the objects with 
  L(5100)$\le 0.5\ 10^{44}$ erg/s, and
 computed according to the self-gravitating 
 disc model with a
  viscosity parameter 
 $\alpha=0.01$. NLS1s are indicated as red dots.}
 \label{fig-dotM-vs-M}
\end{figure}

 There are several causes of uncertainties in the results (cf. Krolik 
 2001 and C02), which might introduce 
 errors on the BH masses as large as one order of magnitude,  
 because one should not forget that even the masses determined directly
with reverberation mapping are  known with an 
uncertainty of a factor 3.  
  It seems however implausible 
  that all the uncertainties would systematically act towards an
  underestimation of the 
  mass and an overestimation of the luminosity,
   avoiding the conclusion of super-Eddington accretion rates.
 Only the uncertainty on the correcting factor of the FWHM due to the 
 geometry and kinematics of the BLR could lead
  to a {\it systematic} underestimation of the mass, if the BLR
   is a rotating flat 
  structure. It can be large when the objects are seen almost face-on. 
  We shall discuss this point in the last 
  section.
   
Fig. \ref{fig-RsgsRg-Tsg01} displays  $R_{\rm sg}/R_{\rm G}$ versus $M$ 
   for all samples, for the self gravitating disc with $\alpha=0.01$.
    We note that
    it is always quite 
   small (in particular smaller than the BLR, which has typical 
   values 10$^{3}$ for high BH masses and 10$^{5}$ for NLS1s), justifying our previous 
  claim that the BLR is always located in, or above, the 
   unstable part of the disc.  As expected $R_{\rm sg}/R_{\rm G}$ decreases with 
   the BH mass, except at the high mass limit, and there is a strong correlation between 
   the two parameters. 
   
 \begin{figure}
 \centering
 \includegraphics[width=9cm]{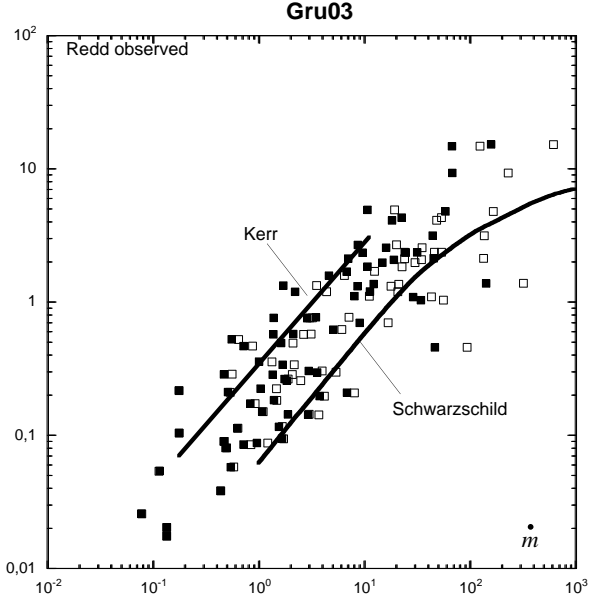}
\caption{The Eddington ratio  $R{\rm edd}$ as a function of $\dot{m}$ for the Gru03 
sample, 
computed with the standard disc (open squares) and the self-gravitating 
disc,
 $\alpha=0.01$ (filled squares).  The 
 objects with L(5100) $\le 0.5\ 10^{44}$ ergs/sec have been suppressed.
 The two curves correspond 
to the slim disc model, $\alpha=0.1$, and respectively a Schwarzshild 
and a kerr BH.} 
 \label{fig-gru-Redd-vs-dotm}
\end{figure}
   
Although the choice of parameters for the self-gravitating disc 
does not influence strongly $\dot{m}$, it has an effect on the optical 
spectral index. As an illustration, Fig. \ref{fig-bor-alfopt-vs-FWHM} 
shows the computed optical spectral 
index $\alpha_{\rm opt}$ defined as $F_{\nu} \propto \nu^{-\alpha_{\rm 
opt}}$ between 4400 and 7000\AA$\ $(rest frame), 
 for the Bor03 sample. The computation is 
 performed with the 
 self-gravitating correction, for 
  a viscosity parameter $\alpha$ equal to 0.01, 0.1, and 0.3. 
  A systematic correction 
 $E(B-V)$=0.05 for the 
  galactic absorption has been applied (certainly an underestimation). 
	For $\alpha=0.01$, the continuum is  
red except for very broad line objects. The trend that broader 
objects have
	bluer optical spectra is consistent with the observational results 
	of Constantin \& Shields (2003). The continuum is globally bluer for smaller 
values of $\alpha$ (0.1 and 0.3). We also see 
that $\alpha_{\rm opt}$ almost never reaches the value of the standard disc 
(-0.33). A detailed comparison with the observed 
values is postponed to the next paper.

Fig. \ref{fig-gru99-alfopt-vs-FWHM} shows a comparison between 
 the observed and computed spectral indexes for the 
Gru99 sample, excluding the 
 objects with L(5100)$\le 0.5\ 10^{44}$ erg/s.  According to Grupe et al. (1999), 
  the observed values of $\alpha_{\rm 
  opt}$ are given with an uncertainty of $\pm 0.4$.
   With $\alpha=0.1$, many of the computed indices are close to the 
  value of the standard disc, while the objects of the samples 
  are particularly red, with an average index of 0.8. The agreement 
  is much better for the smallest viscosity 
  parameter $\alpha=0.01$. The very red 
  spectra observed in a fraction of objects might be due to intrinsic 
  reddening not taken 
  into account in the computed values. If it is the case, it would imply 
  that the observed L(5100) is 
  underestimated in these objects, but again it is not important for 
  the determination of $\dot{m}$. Note that in this sample, NLS1s do not seem to have 
  redder continua than BLR1s. 
  
  It is therefore impossible from this comparison to decide which are 
  the best values of $\alpha$ and $\gamma$ to choose for the disc. Our model is clearly 
  oversimplified, and would require a more sophisticated 
  parametrization. The only conclusion which can be drawn is that a 
  non-standard disc with an additional release of energy in its external 
 region gives a better fit to the average optical 
  continuum of AGN than a standard disc. However, this problem  does not question 
  the existence of super-Eddington accretion 
  rates for NLS1s.
  
  Finally Fig. \ref{fig-dotM-vs-M}
  displays the accretion rates in 
  M$_{\odot}$/yr for the four samples, excluding the objects with 
  L(5100)$\le 0.5\ 10^{44}$ erg/s,  and
 computed according to the self-gravitating 
 disc model with a
  viscosity parameter 
 $\alpha=0.01$. Note that the imposed limit 
 on L(5100) creates the sharp limitation on the left side, as 
 $\dot{M}$ is proportional to $M^{-2/3}$ (for a fixed $L_{\rm opt}$). 
 The limitation on the right 
 side is due to a limitation of $\dot{m}$ at about 0.03 (perhaps due 
 to the fact that the accretion disc changes into an ADAF below this 
 value). NLS1s are indicated as red dots. 
 Despite the large values of $\dot{m}$ 
  of NLS1s, we see that {\it the maximum accretion rate is of the order of one 
  M$_{\odot}$/yr whatever the BH mass}. {\it This is a strong 
  indication of an exterior regulation of the accretion}, 
  rather than the self-regulation of the disc. 
   Note that it is a modest value when compared with the 
  rate of star formation in a starburst nucleus.
   
   \subsection{Comparison with the slim disc model}
   
   It is interesting to compare the observed SED of   
super-Eddington objects with the slim disc model. As we 
mentioned in the introduction, this was done in detail for the two
highest $\dot{m}$ objects (Kawaguchi 2003; KPH; Kawaguchi, Matsumoto,
	Leighly in preparation; see Kawaguchi 2004).
and it will be performed for the objects of the samples in a 
future paper. Here we simply compute
the bolometric luminosity,
and we compare it with the observed values. 

Only the Gru03 sample provides bolometric luminosities based on the observed SEDs.
Fig. \ref{fig-gru-Redd-vs-dotm} shows the observed ratio $R_{\rm 
Edd}$ versus $\dot{m}$ for this sample.
 The  
low luminosity nuclei (L(5100)$\le
 0.5\ 10^{44}$ ergs/sec) have been suppressed. Also shown is the 
 theoretical curves obtained for the slim disc model with a Schwarzschild 
 and a Kerr
 BH. These curves depend very little on the BH mass and on the 
 viscosity parameter. In spite of the large dispersion of the 
 ``observations", it is clear that a majority of points lie above the Schwarzschild
 curve, meaning that the efficiency of the Schwarzschild BH is 
 insufficient, i.e. {\it a Kerr BH with an efficiency of about 0.15 would better 
 fit the observations} unless there is a systematic underestimation of 
 the BH masses. On the other hand, the shape of the curve 
agrees well with the observed points, in particular in the 
 ``saturation" of $R_{\rm Edd}$ above $\dot{m}=10$. Three  objects 
 reach an Eddington ratio of the order of 10, for $50 \le \dot{m}\le 1000$.
 
 \section{Influence of the inclination on the masses and accretion 
 rates}
 
 \begin{figure}
 \centering
 \includegraphics[width=7cm]{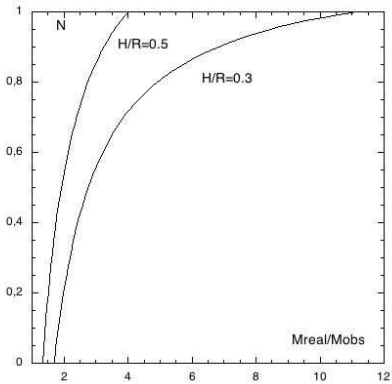}
\caption{Cumulative number 
of objects (normalized to unity) on which an underestimation of the 
mass by a factor 
smaller than 
$M_{\rm real}/M_{\rm obs}$ is made, in the conditions explained in the text. } 
 \label{fig-N-vs-G}
\end{figure}
 
  In all mass determinations, the FWHM is used instead of the dispersion 
 velocity. It makes the implicit assumption that the velocities are 
 distributed at random in the BLR. However, if the BLR is a flat structure
  dominated by rotation, 
 the FWHM is proportional to sin$(i)V_{\rm Kep}$, where $i$ is the 
 angle between the normale and the line of sight (the inclination).
  It is clear that 
 a small inclination can lead to a large underestimation of $V_{\rm Kep}$ and 
 therefore of the mass.
 
 However the BLR cannot be a geometrically thin disc with an exactly 
 Keplerian 
 velocity. Unfortunately its dynamics and its 
 structure are still not well determined from detailed reverberation 
 mappings, but we know that it should be at least a ``thick disc",
  with an aspect ratio 
 larger than, say, $H/R\sim 0.3$ ($H$ being the disc thickness at the 
 radius $R$), since it needs to have a large coverage 
 factor of 
 the central source. Such a disc must be sustained vertically by a turbulent 
 pressure corresponding to a turbulent velocity of the order of 
 $V_{\rm Kep}H/R$. The FWHM is then proportional to
$V_{\rm Kep}\sqrt{(H/R)^2+{\rm sin}(i)^{2}}$,
and the ratio $G$ between the real 
 mass and the ``observed" mass, is: 
 
  \begin{equation}
G=M_{\rm real}/M_{\rm obs}= 
1/[(H/R)^2+{\rm sin}(i)^{2}].
\label{eq-G}
\end{equation} 
 
 We can compute how many objects have a mass underestimated by a given 
 factor $G$, assuming that they are distributed at random inside 
 an angle $i_{0}$. We choose $i_{0}=\pi/4$ in the following 
 computations, as it is a commonly
  accepted value for the opening 
 angle of the dusty torus in 
 Seyfert 1 (according to the Unified Scheme, cf. Antonucci \& Miller 1985). 
  The probability of seeing an object at an inclination angle 
 $i$ per unit angle interval is thus sin$(i)/$[cos$(i_{0})-1]$.  The 
 number of objects per unit interval of $G$ is:
 
 \begin{equation}
{dN\over dG}={[(H/R)^2+{\rm sin}(i)^{2}]^{2}\over 2[1-{\rm cos}(i_0)]
 {\rm cos}(i)}.
\label{eq-dNdG}
\end{equation}
 
  Fig. \ref{fig-N-vs-G} shows the integral of this expression, i.e. 
  the cumulative number 
of objects (normalized to unity) on which an underestimation of the 
mass by a factor  
smaller than
$M_{\rm real}/M_{\rm obs}$ is made, for 
  two values of $H/R$. Note that $i_{0}$ plays
  a non negligible role here, as it contributes to increase the 
  proportion of objects having a large $G$ by a factor 3 with 
  respect to an isotropic distribution.
  We see that the factor $G$ depends strongly 
  on the aspect ratio the BLR. For $H/R=0.3$, it can take values as 
  large as 11, but the number of objects reaching this value is 
  small: only about 20\% have a $G$-factor larger than 6, and 1\% a 
  factor larger than 10. For $H/R=0.5$, the maximum value of $G$ is only 
  4, and about 60\% have a $G$-factor of the order of or smaller than 2. 
  
  It is important to realize that in this case {\it not only  nearly 
  face-on objects, but all Seyfert 1} will have their BH mass underestimated. 
  This would raise a problem concerning the BH-bulge mass 
  relationship. 
  
 Would we have thus to modify our conclusions? For 
   $H/R=0.3$, half of 
  NLS1s could have their masses underestimated by factors of 3 to 10, 
  leading to underestimations of the accretion rates (in terms of 
  Eddington) by factors 10 to 100. It is clearly very important, but 
   still a large proportion of objects would be accreting 
  above the Eddington limit, however at a smaller rate. On the other hand, 
  it is quite possible that the few extremely high accretors are 
  actually ``face-on" objects, and that their mass is indeed underestimated 
  by about one order of magnitude.

\section{Conclusion}
     
     We used the BH masses deduced from 
     the size-luminosity relationship to compute their accretion rate in four 
     samples of AGN, assuming that the optical luminosity is provided 
     by the accretion disc. Thus the empirical relation 
     must be extrapolated in a range of masses almost one order of 
     magnitude 
     smaller than the Kaspi et al. sample. We used a simplified disc model, 
      with a parametrization of the energy release in the self-gravitating 
     region to get the accretion rate, and the slim disc model in the 
     inner regions in order to get the bolometric luminosity.
       In spite of the crudeness of the treatment, 
     this study leads to several fairly certain
      conclusions.
     
 \begin{itemize}
 \item NLS1s are always accreting at Eddington or super-Eddington 
 rates. $\dot{m}$ can reach 1000, corresponding to an 
 accretion rate equal to 60$\dot{M}_{\rm Edd}$ (for Schwarzschild BH) and 
 to 300$\dot{M}_{\rm Edd}$ (for Kerr BH). 
 \item Their observed bolometric luminosities ``saturate" at $\sim$ 10 Eddington 
 luminosities, as predicted by slim disc models. It explains why there 
 is a lower limit to the 
 observed
 FWHM.
 \item The observed value of the bolometric luminosities are in better agreement 
 with a Kerr than with a Schwarzschild BH.
 \item  The computed optical spectral indexes agree with 
 the observed trend of redder spectra for NLS1s than for BLS1s.
 \item And finally the accretion rates have an upper limit 
 of about one 
  M$_{\odot}$/yr, whatever the BH mass. In particular, all NLS1s 
  have an accretion rate of this order. This is a strong 
  indication for a mass limited supply, implying
   an exterior regulation of the accretion. 
  \end{itemize}

With these results we are in a position to say now that NLS1s should have a strong 
influence on 
 the growth of BHs.  This is in agreement with the claim by
 Mathur et al. 2001, and Grupe \& 
Mathur 2004. Since NLS1s constitute about 10$\%$ 
 of normal Seyfert 
which themselves are about 2$\%$ of inactive galaxies, one 
deduces that all galaxies spend 0.2$\%$ of their lifetime 
in the NLS1 phase, i.e. 2 
10$^{7}$ years. During this time the mass of the BH increases by one 
order of magnitude (Kawaguchi et al. 2004). 
This could account  both for the observed large dispersion in the 
BH/bulge mass relation of NLS1s, and for the existence of undermassive 
BH/bulge ratios
during a large fraction of the NLS1 phase. The increase of the bulge 
 mass could have taken place during merger or 
 interaction events. BHs would then grow 
 during intense phases of activity after a time delay, necessary for 
 accumulating matter in the circumnuclear region and for
 triggering a starburst. 
In this scenario, the overabundance of iron could be easily explained
by the rapid formation of massive stars and supernovae explosions in the 
outer parts of the accretion disc where the accretion rate is high (Collin \& Zahn 2000, 
Levin 2003, Levin \& Belobodorov 2003).
The scenario would also
account naturally for the presence of outflows giving 
rise to the blue wing of the [OIII] line, as super-Eddington accretion 
is expected to generate outflows by strong radiation fields.

\medskip 

 Though we have tried to determine a lower limit of the accretion rate, 
two effects can intervene to still reduce it. 
They were both discussed 
in C02. 
\medskip

\noindent 1- the possibility that the accretion rate decreases with the 
radius between the optically emitting region and the BH, owing to the 
creation of a strong outflow due to the 
 radiation pressure. The accretion rate close to the BH would 
then be just Eddington. In this case, the  outflow could
 well be the origin of the [OIII] wing, and could lead to the escape of a part
  of the Narrow Line Region, explaining the weakness of the 
  [OIII] line. However, one should 
realize that in this case the rate of outflow would have to represent 90 or 
even 99$\% $ of the accretion rate, in the highest accretors.  
This seems unrealistic.

\medskip

\noindent 2-  The 
optical luminosity is not provided by the accretion disc. Recently 
King \& Pounds (2003) suggested that BHs accreting at super Eddington 
rate produce winds which are Thomson thick and can emit a black 
body spectrum providing the Blue Bump of AGN. 
Pounds et al. (2003) indeed report that they have found the signature of such 
an optically thick wind in the X-ray spectrum of the NLS1 PG1211+143. 
If the existence of such a wind is confirmed in other NLS1s, then it 
is clear that the present analysis would have to be reconsidered. 
However let us recall that Collin et al. (2002) have shown 
that very strong conditions must be met in such a wind to give 
rise to the optical-UV featureless continuum: it must
have both a large density (10$^{14}$ 
cm$^{-3}$), and a Thomson thickness at least of 
unity. Besides, to get the observed luminosity,
it should be located far 
from the center and it should have a large spatial extension. It is 
thus not obvious that the wind
 observed by Pounds et al. 
(2003) satisfies these requirements. It is more likely that 
its emission is limited 
only to the EUV radiation, and that  
the optical emission is still due to the accretion disc. 
\bigskip

 We have assumed all along the paper 
that the BH masses of NLS1s are correctly estimated by 
the empirical reverberation relations, even when these relations had 
to be extrapolated by almost one order of magnitude. On the other 
hand, we have accepted the usual assumption that the FWHM is a good 
measure of the velocity in the BLR, implicitly assuming that the 
velocities are distributed at random. If on the contrary the BLR is a 
flat structure dominated by rotation, the BH masses of a fraction of 
objects could be 
underestimated by factors up to one order of magnitude and the 
accretion rates (in terms of Eddington) by two order of magnitudes 
when they are seen nearly face-on. 
However, since this 
fraction should be small, we think that the scenario described 
in this paper is qualitatively correct.  

\begin{acknowledgements}
We are grateful to Amri Wandel for useful comments which have 
contributed to 
improve substantially the paper.
\end{acknowledgements}

\bigskip

\end{document}